\documentclass[fleqn,10pt]{wlscirep}

\usepackage{epsfig}
\usepackage{amsmath}
\usepackage{graphicx}
\usepackage{dcolumn}
\usepackage{bm}
\usepackage{epstopdf}
\usepackage{bbold}
\usepackage{amsthm}
\usepackage{subfigure}
\newtheorem{theorem}{Theorem}

\newcommand{\iden}{\mathbb{1}}

\title{Quantum coherence of steered states}

\author[1,*]{Xueyuan Hu}
\author[2]{Antony Milne}
\author[1]{Boyang Zhang}
\author[3]{Heng Fan}
\affil[1]{School of Information Science and Engineering, and Shandong Provincial Key Laboratory of Laser Technology and Application, Shandong University, Jinan, 250100, P. R. China}
\affil[2]{Controlled Quantum Dynamics Theory, Department of Physics, Imperial College London, London SW7 2AZ, UK
}
\affil[3]{Institute of Physics, Chinese Academy of Sciences, Beijing 100190, China}

\affil[*]{xyhu@sdu.edu.cn}

\begin{abstract}
Lying at the heart of quantum mechanics, coherence has recently been studied as a key resource in quantum information theory. Quantum steering, a fundamental notion originally considered by Schr{\"o}dinger, has also recently received much attention. When Alice and Bob share a correlated quantum system, Alice can perform a local measurement to `steer' Bob's reduced state. We introduce the maximal steered coherence as a measure describing the extent to which steering can remotely create coherence; more precisely, we find the maximal coherence of Bob's steered state in the eigenbasis of his original reduced state, where maximization is performed over all positive-operator valued measurements for Alice. We prove that maximal steered coherence vanishes for quantum-classical states whilst reaching a maximum for pure entangled states with full Schmidt rank. Although invariant under local unitary operations, maximal steered coherence may be increased when Bob performs a channel. For a two-qubit state we find that Bob's channel can increase maximal steered coherence if and only if it is neither unital nor semi-classical, which coincides with the condition for increasing discord. Our results show that the power of steering for coherence generation, though related to discord, is distinct from existing measures of quantum correlation.
\end{abstract}
\begin{document}

\flushbottom
\maketitle
%
%
\thispagestyle{empty}



Quantum coherence, originating from the quantum pure state superposition principle, is one of the most fundamental properties of quantum mechanics. It is increasingly recognized as a vital resource in a range of scenarios, including quantum reference frames~\cite{refframes,Marvian,Marvian2}, transport in biological systems~\cite{bio1, bio2, bio3} and quantum thermodynamics~\cite{thermo1, thermo2, thermo3}. How to measure coherence is an essential problem in both quantum theory and quantum information and has recently attracted much attention \cite{PhysRevLett.113.140401,PhysRevLett.113.170401,PhysRevA.91.042120,PhysRevA.91.042330,PhysRevA.91.052115}. The quantification of coherence in a single quantum system depends on both the quantum state and a fixed basis for the density matrix of the system \cite{PhysRevLett.113.140401,PhysRevLett.113.170401}. The fixed basis is usually chosen as the eigenbasis of the Hamiltonian or another observable. In either case, the quantified coherence is not an intrinsic property of the single-party quantum state itself. The dynamics of quantum coherence under certain noisy channels has also attracted a lot of research attention \cite{Zhang15,Chin} and is connected to the dynamics of quantum correlations \cite{PhysRevLett.114.210401}.

When Alice and Bob share a correlated quantum system, a measurement by Alice can `steer' the quantum state of Bob. Quantum steering, especially Einstein-Podolsky-Rosen (EPR) steering, has long been noted as a distinct nonlocal quantum effect \cite{steer35} and has attracted recent research interest both theoretically and experimentally \cite{PhysRevLett.98.140402,Nat.Phys.2010,Nat.Photon.2012,PhysRevLett.112.180404}. The quantum steering ellipsoid (QSE) \cite{Verstraete2002,Shi2011,PhysRevLett.113.020402,steer_njp,PhysRevA.92.012311}, defined as the whole set of Bloch vectors to which Bob's qubit can be steered by a positive-operator valued measurement (POVM) on Alice's qubit, provides a faithful geometric presentation for two-qubit states. Using the QSE formalism we have studied a class of two-qubit states whose quantum discord can be increased by local operations \cite{PhysRevA.91.022301}. Interestingly, arbitrarily small mutual information is sufficient for the QSE of a pure two-qubit state to be the whole Bloch ball. Since mutual information is an upper bound of quantum correlation measures such as entanglement and discord, the power that one qubit has to steer another cannot be fully characterized by the quantum correlation between the two qubits. A measure that quantifies the power of generating quantum coherence by steering is therefore necessary.

In this paper we consider a bipartite quantum state $\rho$ with non-degenerate reduced state $\rho_B$ and study the coherence of Bob's steered state, which is obtained by Alice's POVM. Here the eigenbasis of $\rho_B$ is employed as the fixed basis in which to calculate the coherence of the steered state. The significance of this choice of basis is that Bob's initial state is incoherent. When Alice performs a local measurement, she can steer Bob's state to one that is coherent in the eigenbasis of $\rho_B$, i.e. Alice generates Bob's coherence. By $\mathcal C(\rho)$ we denote the maximum coherence that Alice can generate through local measurement and classical communication. In contrast to existing quantifiers of coherence, $\mathcal C$ is an intrinsic property of the bipartite quantum state $\rho$, because the reference basis of coherence, chosen as the eigenbasis of $\rho_B$, is inherent to the bipartite state. Furthermore, we find that $\mathcal C$ gives a different ordering of states compared to quantum entanglement or discord; this indicates that $\mathcal C$ describes remote quantum properties distinct from these measures of quantum correlation. Properties of $\mathcal C$ are also studied. The maximal steered coherence is found to vanish only for classical states and can be created and increased by local quantum channels. Given that coherence plays a central role in a diverse range of quantum information processing tasks, we can also consider how steered coherence might be used as a resource. We close our discussion by presenting one such scenario.

We note that, shortly after this paper first appeared, Mondal \emph{et al.} presented a study on the steerability of local quantum coherence~\cite{Mondal}. We consider our works to be complementary: though examining a similar topic, our approaches are very different (Mondal \emph{et al.} consider steering from the existence of a local hidden state model rather than from the perspective of the QSE formalism).

\section*{Results}

\subsection*{Definition.}
We consider a bipartite quantum state $\rho$, where the reduced state $\rho_B$ is non-degenerate with eigenstates $\Xi=\{|\xi_i\rangle\}$. When Alice obtains the POVM element $M$ as a measurement outcome, Bob's state is steered to \mbox{$\rho_B^M:=\mathrm{tr}_A(M\otimes \mathbb{1}\,\rho)/p_M$} with probability $p_M:=\mathrm{tr}(M\otimes \mathbb{1}\,\rho)$, where $\mathbb{1}$ denotes the single qubit identity operator. Baumgratz \emph{et. al.}\cite{PhysRevLett.113.140401} gives the the quantum coherence $C$ of $\rho_B^M$ in the basis $\{|\xi_i\rangle\}$ as the summation of the absolute values of off-diagonal elements:
\begin{equation}
C(\rho_B^M,\{|\xi_i\rangle\})=\frac{1}{p_M}\sum_{i\neq j}\big|\langle\xi_i|\mathrm{tr}_A(M\otimes \mathbb{1}\,\rho)|\xi_j\rangle\big|.
\end{equation}
Here we maximize the coherence $C(\rho_B^M,\{|\xi_i\rangle\})$ over all possible POVM operators $M$ and define the maximal steered coherence as
\begin{equation}
\mathcal C(\rho):=\max_{M\in\mathrm{POVM}}\left[\frac{1}{p_M}\sum_{i\neq j}\big|\langle\xi_i|\mathrm{tr}_A(M\otimes \mathbb{1}\,\rho)|\xi_j\rangle\big|\right].\label{coherence}
\end{equation}
When $\rho_B$ is degenerate, $\Xi$ is not uniquely defined; however, we can take the infimum over all possible eigenbases for Bob and define the maximal steered coherence as
\begin{equation}
\mathcal C(\rho):=\inf_\Xi\left\{\max_{M\in\mathrm{POVM}}\left[\frac{1}{p_M}\sum_{i\neq j}\big|\langle\xi_i|\mathrm{tr}_A(M\otimes \mathbb{1}\,\rho)|\xi_j\rangle\big|\right]\right\}.\label{coherence_d}
\end{equation}

It is worth noting that $\mathcal C$ is an intrinsic property of the bipartite quantum state $\rho$. When fixing the basis in which to calculate the coherence, we need not choose an observable that is independent of the state; the basis $\{|\xi_i\rangle\}$ we choose here is inherent to the state $\rho$.

\subsection*{Properties.} We prove that the following important properties hold for maximal steered coherence.

(E1) $\mathcal C$ vanishes if and only if $\rho$ is a classical state (zero discord for Bob), i.e. can be written as
\begin{equation}
\rho=\sum_{i=1}^{d}p_i\rho_i^A\otimes|\xi_i\rangle\langle\xi_i|.\label{classical}
\end{equation}
The proof of this is given in Methods.

(E2) $\mathcal C$ reaches a maximum for all pure entangled states with full Schmidt rank, i.e. states that can be written in as $|\Psi\rangle=\sum_{i=1}^{d_B}\lambda_i|\phi_i^A\rangle\otimes|\xi_i^B\rangle$ with $\lambda_i\neq0\,\,\forall i$. Here $d_B$ is the dimension of Bob's state. For a single quantum system of dimension $d_B$, the maximally coherent state is $|\psi_m^B\rangle=\frac{1}{\sqrt{d_B}}\sum_{i=1}^{d_B}|xi_i^B\rangle$ \cite{PhysRevLett.113.140401}; Bob is steered to this when Alice obtains the measurement outcome $M=|\psi^A\rangle\langle\psi^A|$, where $|\psi^A\rangle$ is the state $\sum_{i=1}^{d_B}\frac{1}{\lambda_i}|\phi_i^A\rangle$ after normalisation.

(E3) $\mathcal C$ is invariant under local unitary operations. When the unitary operator $U=U_A\otimes U_B$ acts on a bipartite state $\rho$, the eigenbasis of $\rho_B$ is rotated by $U_B$, so that the off-diagonal elements of $\rho_B^M$ become
\begin{eqnarray}
\langle\xi_i|U^\dagger_B\mathrm{tr}_A[M\otimes \iden\,(U\rho U^\dagger)]U_B|\xi_j\rangle=\langle\xi_i|\mathrm{tr}_A[U_A^\dagger MU_A\otimes \iden\,\rho ]|\xi_j\rangle.
\end{eqnarray}
From Eq. (\ref{coherence}) it is clear that $\mathcal C(U\rho U^\dagger)=\mathcal C(\rho)$.

(E4) $\mathcal C$ can be increased by Bob performing a local quantum channel prior to Alice's steering.

Property (E4) holds owing to the fact that a local channel $\Lambda_B$, under certain conditions \cite{PhysRevA.85.032102}, can transform a classical state with vanishing $\mathcal C$ into a discordant state with strictly positive $\mathcal C$. Note, however, that a channel $\Lambda_A$ performed by Alice prior to steering cannot increase $\mathcal C$. (This follows because $\Lambda_A$ can be performed by applying a unitary operation to $A$ and an ancilla $A'$ and then discarding $A'$; the unitary operation does not affect the set of Bob's steered states, while discarding $A'$ may limit Alice's ability to steer Bob's state. Thus $\Lambda_A$ performed by Alice does not alter Bob's reduced state $\rho_B$ but shrinks the set of his steered states; such a channel cannot increase $\mathcal C$.)

Let us also note an important consequence of property (E2): $\mathcal C$ is distinct from the entanglement $E$ \cite{PhysRevLett.80.2245} and discord-type quantum correlations $\mathcal D$ \cite{PhysRevA.67.012320}. In fact, $\mathcal C$ gives a different ordering of states from $E$ or $\mathcal D$. We demonstrate this by considering states $\rho_1=|\Psi_1\rangle\langle\Psi_1|$ and $\rho_2=(1-\delta)|\Psi_2\rangle\langle\Psi_2|+\delta\frac{\iden_d\otimes\iden_d}{d^2}$, where $|\Psi_1\rangle=\sqrt{1-(d-1)\delta}|00\rangle+\sqrt{\delta}\sum_{j=1}^{d-1}|jj\rangle$ and $|\Psi_2\rangle=\frac{1}{\sqrt d}\sum_{j=0}^{d-1}|jj\rangle$ are both pure entangled states with full Schmidt rank of dimension $d$, and $0<\delta\ll1$. According to (E2), $\mathcal C(\rho_1)$ reaches the maximum; whereas for $\rho_2$, Bob's steered state is always mixed and hence not maximally coherent state in any given basis. We therefore have $\mathcal C(\rho_1)>\mathcal C(\rho_2)$. Meanwhile, $\mathcal D(\rho_1)$ and $E(\rho_1)$ can be made arbitrarily small by taking $\delta$ to be small enough, whilst $\mathcal D(\rho_2)$ and $E(\rho_2)$ approach 1 for small $\delta$. Hence $\delta$ exists such that $\mathcal D(\rho_1)<\mathcal D(\rho_2)$ and $E(\rho_1)<E(\rho_2)$.

\subsection*{General expression for two-qubit states.} We now derive the general form of $\mathcal C$ for two-qubit states. The state of a single qubit can be written as $\rho_B=\frac12\sum_{i=0}^3b_i\sigma_i$, where $\sigma_0=\iden$, $\sigma_i$ with $i=1,2,3$ are Pauli matrices, $b_i=\mathrm{tr}(\rho_B\sigma_i)$, and $\boldsymbol b=(b_1,b_2,b_3)^\mathrm T$ is Bob's Bloch vector. The norm of the vector $\boldsymbol b$ is denoted by $b$. The quantum coherence of $\rho_B$ in a given basis $\{\boldsymbol n,-\boldsymbol n\}$, where $|\boldsymbol n|=1$, is
\begin{equation}
C(\rho_B,\boldsymbol n)=\big|\boldsymbol b\times\boldsymbol n\big|.\label{coherence1}
\end{equation}
Let $B$ and $N$ be the points associated with the vectors $\boldsymbol b$ and $\boldsymbol n$ respectively, and let $O$ be the origin. Since $\frac12\big|\boldsymbol b\times\boldsymbol n\big|$ is the area of $\triangle{OBN}$ and the line segment $\overline{ON}$ is unit length, $C(\rho_B,\boldsymbol n)$ is simply the perpendicular distance between the point $B$ and the line $\overline{ON}$.

Similarly, we can write a two-qubit state in the Pauli basis as \mbox{$\rho=\frac14\sum_{i,j=0}^3\Theta_{ij}\,\sigma_i^A\otimes\sigma_j^B$}, where the coefficients $\Theta_{ij}=\mathrm{tr}(\rho\,\sigma_i^A\otimes\sigma_j^B)$ form a block matrix $\Theta=\left(\begin{array}{cc}1&\boldsymbol b^\mathrm T\\ \boldsymbol a & T\end{array}\right)$. Here $\boldsymbol a$ and $\boldsymbol b$ are Alice and Bob's Bloch vectors respectively, and $T$ is a $3\times3$ matrix. Note that when $\rho_B$ is non-degenerate we have $\boldsymbol b\neq \boldsymbol 0$. We ignore the trivial case that $a=1$, when $\rho_A$ is pure and hence $\rho$ is a product state.

When the POVM operator $M=\frac12(\sigma_0+\boldsymbol m\cdot\boldsymbol\sigma)$ is obtained on Alice's qubit, Bob's state becomes
\begin{equation}
\boldsymbol b_M=\frac{\boldsymbol b+T^\mathrm T\boldsymbol m}{|1+\boldsymbol a\cdot\boldsymbol m|}.
\end{equation}
Here $\boldsymbol\sigma=(\sigma_1,\sigma_2,\sigma_3)^\mathrm T$, and $\boldsymbol m=(m_1,m_2,m_3)^\mathrm T$ can be any point on or inside the Bloch sphere. The set of $\boldsymbol b_M$ forms the QSE $\mathcal E_B$. When $\rho_B$ is non-degenerate, we have $\boldsymbol b\neq\boldsymbol 0$. According to Eq. (\ref{coherence1}), the coherence of $\boldsymbol b_M$ in the basis $\{|\xi_i\rangle\}$ is \mbox{$C(\rho_B^M,\boldsymbol n_B)=\big|\boldsymbol b_M\times\boldsymbol n_B\big|$}, with $\boldsymbol n_B=\boldsymbol b/b$; this represents the perpendicular distance from the point $B_M$ to the line $\overline{OB}$ (Figure \ref{fig:examples}\textbf{a}). Hence the maximal steered coherence $\mathcal C(\rho)$, as defined in Eq. (\ref{coherence}), is the maximal perpendicular distance between a point on the surface of $\mathcal E_B$ and $\overline{OB}$. Explicitly, we have $C(\rho_B^M,\boldsymbol n_B)=|(T^\mathrm T\boldsymbol m)\times\boldsymbol n_B|\big/|1+\boldsymbol a\cdot\boldsymbol m|$ and
\begin{equation}\label{C_eqn}
\mathcal C(\rho)=\max_{\boldsymbol m\in{\mathbb R}^3, m=1}\bigg|\frac{T^\mathrm T\boldsymbol m\times\boldsymbol n_B}{1+\boldsymbol a\cdot\boldsymbol m}\bigg|.
\end{equation}
The maximization needs to be performed only over all projective measurements with $m=1$ because steered states on the surface of $\mathcal E_B$ correspond to measurements $\boldsymbol m$ on the surface of the Bloch sphere.

When $\rho_B$ is degenerate, $\boldsymbol b=\boldsymbol 0$ and $\boldsymbol n_B$ is arbitrary; the infimum can then be taken over all $\boldsymbol n_B$ to give the maximal steered coherence of a two-qubit state as
\begin{equation}\label{C_eqn_d}
\mathcal C(\rho)=\inf_{\boldsymbol n_B\in{\mathbb R}^3, n_B=1}\left(\max_{\boldsymbol m\in{\mathbb R}^3, m=1}\bigg|\frac{T^\mathrm T\boldsymbol m\times\boldsymbol n_B}{1+\boldsymbol a\cdot\boldsymbol m}\bigg|\right).
\end{equation}

\subsection*{Properties for two-qubit states.}
We now study two-qubit states in more detail; this allows us to identify some important features of the maximal steered coherence, as well as giving a clear geometric interpretation of $\mathcal C$ using the steering ellipsoid formalism.

\begin{figure}[ht]
\centering
\includegraphics[width=0.8\linewidth]{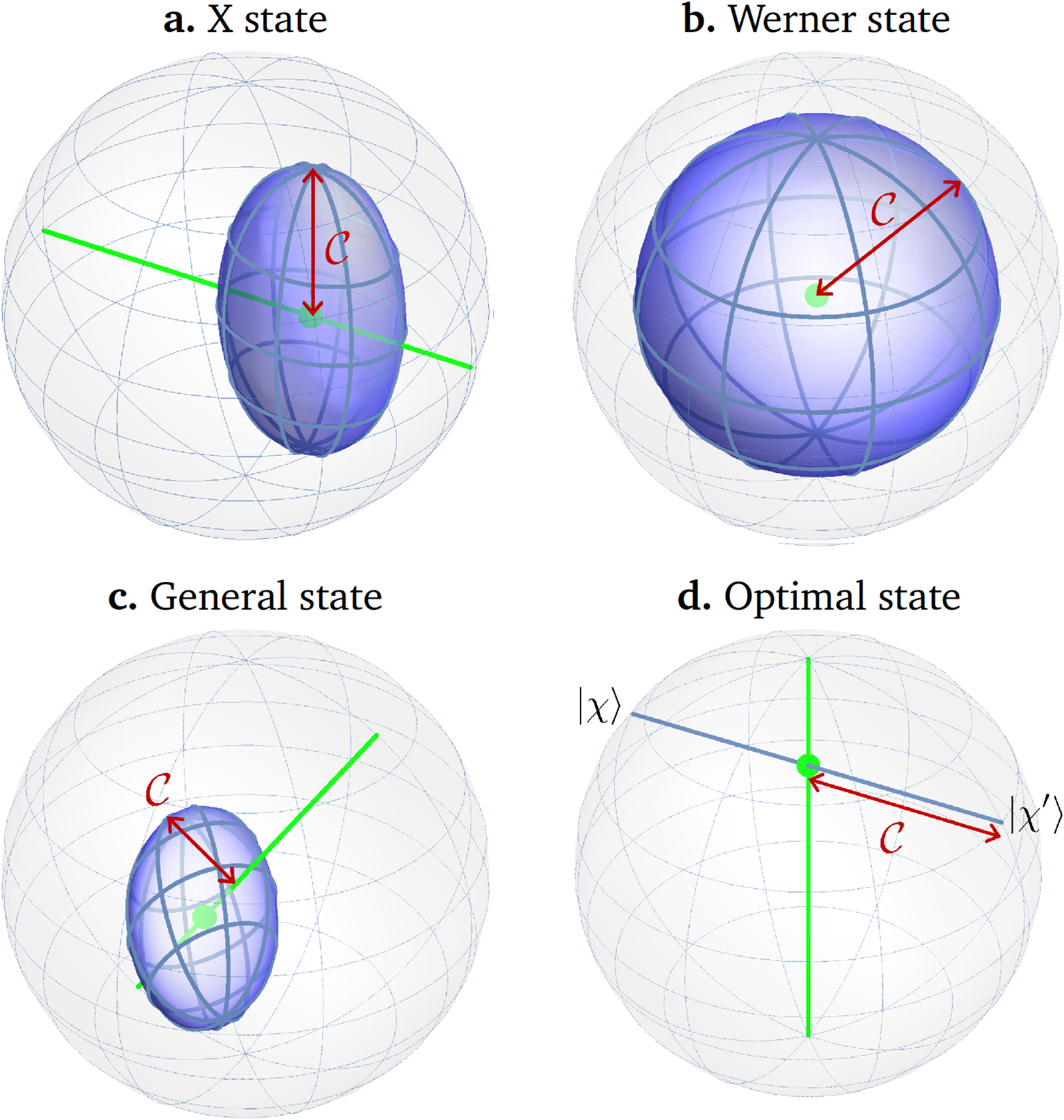}\\
\caption{An illustration of the geometric interpretation of maximal steered coherence $\mathcal C(\rho)$ for two-qubit states $\rho$ using the QSE $\mathcal E_B$. For simplicity we take $\boldsymbol a = \boldsymbol 0$. The point $B$ representing Bob's Bloch vector is indicated by a green blob, and the line $\overline{OB}$ is also shown in green; states lying along this line are incoherent in the basis $\rho_B$. $\mathcal C(\rho)$ is given by the maximal perpendicular distance between a point on $\mathcal E_B$ and $\overline{OB}$; this is shown by the red arrow. \textbf{a} Theorem \ref{thm:canonical} shows that for any canonical state, $\mathcal C(\rho)$ is bounded by the longest semiaxis of the QSE. \textbf{b} A state of the form \eqref{dlc1}, which achieves maximal $\mathcal C(\rho)$ for a given $b$. The QSE is a chord perpendicular to $\overline{OB}$. \textbf{c} When $\rho$ is an X state, $\overline{OB}$ lies along an axis of the QSE, and $\mathcal C(\rho)$ is the length of the longest of the other two semiaxes. \textbf{d} When $\rho$ is a Werner state, $\mathcal E_B$ is a ball centred on the origin. In this case, even though $\rho_B$ is degenerate, $\mathcal C(\rho)$ is well-defined as the radius of the ball.}
\label{fig:examples}
\end{figure}

As demonstrated by property (E4), a trace-preserving channel $\Lambda_B$ performed by Bob may increase $\mathcal C$; we now study an explicit example. Say that Alice and Bob share the classical two-qubit state
\begin{equation}
\rho_c=t|++\rangle\langle++|+(1-t)|--\rangle\langle--|,
\end{equation}
with $\frac12<t<1$ and $|\pm\rangle=\frac{1}{\sqrt 2}(|0\rangle\pm|1\rangle)$. When Bob applies the single-qubit amplitude damping channel, the state transforms as $\rho_c'=\iden_A\otimes\Lambda_B^\mathrm{AD}(\rho_c)$, where \mbox{$\Lambda^\mathrm{AD}(\cdot)=\sum_{i}E_i(\cdot)E_i^\dagger$} with $E_0=|0\rangle\langle0|+\sqrt{1-\gamma}|1\rangle\langle1|$ and $E_1=\sqrt \gamma|0\rangle\langle1|$. We then find that the maximal steered coherence of the transformed state is
\begin{equation}
\mathcal C(\rho_c')=\frac{2t\gamma\sqrt{1-\gamma}}{\sqrt{(1-2t)^2(1-\gamma)+\gamma^2}}.
\end{equation}
$\mathcal C(\rho_c')$ vanishes when $\gamma=0$, becomes positive for \mbox{$0<\gamma<1$}, and then vanishes again at $\gamma=1$.

Maximal steered coherence can be increased by Bob's local amplitude damping channel even when Alice and Bob share a non-classical state. Consider the two-qubit state
\begin{equation}
\rho_p=p|\Psi\rangle\langle\Psi|+\frac{1-p}{4}\sigma_0\otimes\sigma_0,\label{pi}
\end{equation}
where $0<p<1$ and $|\Psi\rangle=\cos\frac{\theta}{2}|++\rangle+\sin\frac{\theta}{2}|--\rangle$. The QSE for such a state is an ellipsoid centered at $(\frac{p(1-p)\cos\theta}{1-(p\cos\theta)^2},0,0)$ with semiaxes of length $c_1=\frac{p(1-p\cos^2\theta)}{1-(p\cos\theta)^2},c_2=c_3=\frac{p\sin\theta}{\sqrt{1-(p\cos\theta)^2}}$ aligned with the coordinate axes ($\mathcal E_B$ is in fact a prolate spheroid as $c_1>c_2=c_3$). Bob's Bloch vector is $\boldsymbol b=(p\cos\theta,0,0)$, which lies on the $x$ axis. The maximal steered coherence is therefore
\begin{equation}
\mathcal C(\rho_p)=\frac{p\sin\theta}{\sqrt{1-(p\cos\theta)^2}}.
\end{equation}
For $p<\frac{1}{2\sin\theta+1}$, the state $\rho_p$ has zero entanglement but nonzero $\mathcal C$. Note also that $\mathcal C(\rho_p)$ is related to both the fraction of $|\Psi\rangle$ and the entanglement associated with $|\Psi\rangle$.

\begin{figure}[ht]
\centering
\includegraphics[width=0.8\linewidth]{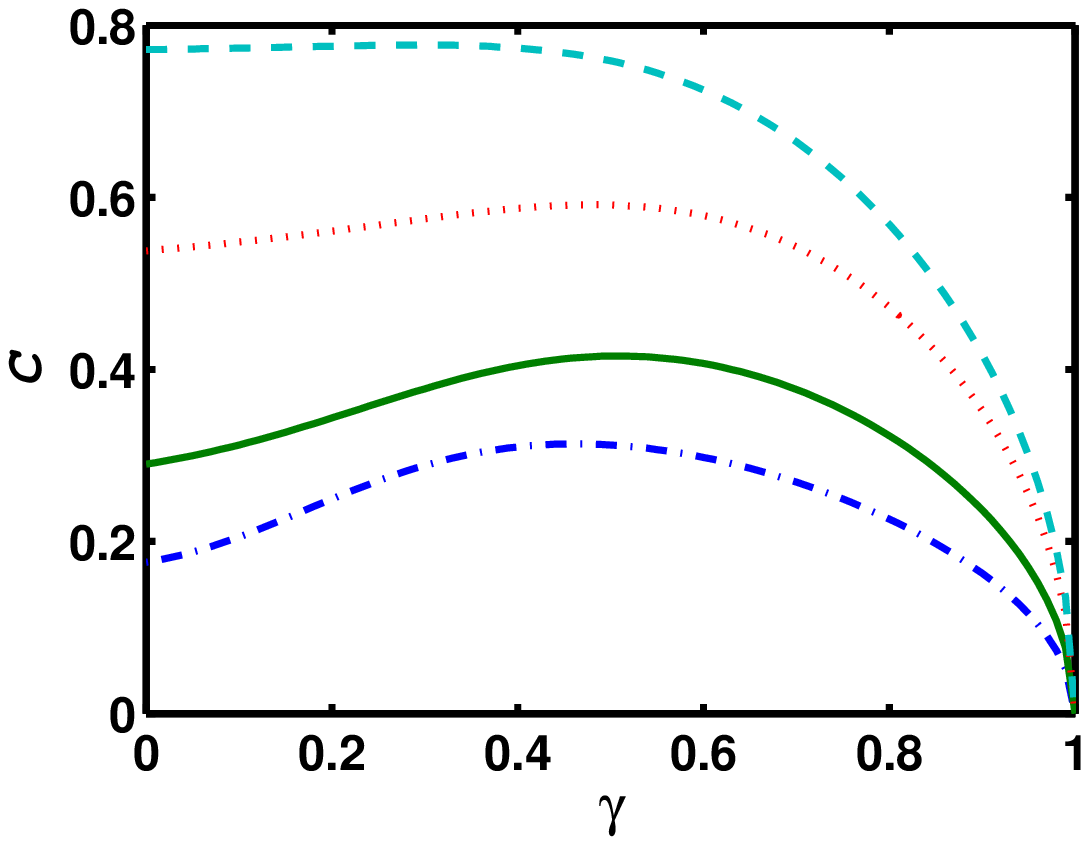}\\
\caption{The evolution of maximal steered coherence under Bob's local amplitude damping channel: $\mathcal C(\iden_A\otimes\Lambda_B^\mathrm{AD}(\rho_p))$, with $\rho_p$ given by Eq. (\ref{pi}). The parameters for the four curves are $p=0.9,\theta=0.2\pi$ for the cyan dashed line; $p=0.9,\theta=0.1\pi$ for the red dotted line; $p=0.7,\theta=0.1\pi$ for the green solid line; and $p=0.5,\theta=0.1\pi$ for the blue dash-dotted line. The corresponding semiaxes ratios, which give a measure of the prolateness of $\mathcal E_B$, are $c_3/c_1=0.980,0.859,0.629\text{ and }0.496$ respectively. The effect of locally increasing  $\mathcal C$ is stronger for more prolate $\mathcal E_B$.}
\label{adc}
\end{figure}

Figure \ref{adc} shows the evolution of $\mathcal C$ under the channel $\Lambda_B^{\mathbb{AD}}$, i.e. $\mathcal C(\rho_p')$, where $\rho_p'=\iden_A\otimes\Lambda_B^\mathrm{AD}(\rho_p)$. By altering $p$ and $\theta$ we alter the ratio of the axes $c_3/c_1$. The results indicate that the potential for increasing $\mathcal C$ under Bob's local amplitude damping is related to the ratio $c_3/c_1$: the smaller the ratio, the stronger the local increase of $\mathcal C$. In other words the effect is strongest when the QSE $\mathcal E_B$ is highly prolate (`baguette-shaped').

In fact, it is possible to formulate a necessary and sufficient condition for the increase of maximal steered coherence for two-qubit states.

\begin{theorem}\label{thm:increase}Bob's local qubit channel $\Lambda_B$ can increase maximal steered coherence for some input two-qubit state if and only if $\Lambda_B$ is neither unital nor semi-classical.
\end{theorem}

The proof is given in Methods. We therefore see that the behavior of maximal steered coherence $\mathcal C$ under local operations is similar to that of quantum discord $\mathcal D$. The set of local channels that can increase $\mathcal C$ for some two-qubit state is the same as the set of local channels that can increase $\mathcal D$. Moreover, $\mathcal C$ can be increased when the QSE $\mathcal E_B$ is very prolate; we showed in Reference \cite{PhysRevA.91.022301} that the quantum discord of Bell-diagonal states with such baguette-shaped $\mathcal E_B$ can be increased by the local amplitude damping channel. We therefore conjecture the local increase in quantum correlations originates from the increase in steered coherence.

We now investigate the set of so-called canonical states, which have particular significance in the steering ellipsoid formalism~\cite{Shi2011,PhysRevLett.113.020402,PhysRevA.90.024302}. Here, a canonical state $\rho_\mathrm{can}$ corresponds to one for which Alice's marginal is maximally mixed ($\boldsymbol a = \boldsymbol 0$). This implies that the QSE $\mathcal E_B$ is centered at $B$ (Figure \ref{fig:examples}\textbf{a}). Let $c_1$, $c_2$ and $c_3$ be the lengths of the semiaxes of $\mathcal E_B$ ordered such that $c_1\geq c_2\geq c_3$.

\begin{theorem}\label{thm:canonical} For any canonical state $\rho_\mathrm{can}$ the maximal steered coherence is bounded by the longest semiaxis. This in turn is bounded by the length of Bob's Bloch vector as
\begin{equation}\mathcal C(\rho_\mathrm{can})\leq c_1\leq \sqrt {1-b^2}.\end{equation}
The bound is saturated if and only if $\mathcal E_B$ is a chord perpendicular to $\boldsymbol b$ meeting the surface of the Bloch sphere at $|\chi\rangle$ and $|\chi'\rangle$. This represents a canonical state of the form
\begin{equation}
\rho_\mathrm{can}=\frac12|\psi\rangle\langle\psi|\otimes|\chi\rangle\langle\chi|+\frac12|\bar\psi\rangle\langle\bar\psi|\otimes|\chi'\rangle\langle\chi'|,\label{dlc1}
\end{equation}
where $\langle\psi|\bar\psi\rangle=0$.
\end{theorem}

The proof is given in Methods, and an example QSE for an optimal state of the form (\ref{dlc1}) is shown in Figure \ref{fig:examples}\textbf{b}). Note that this bound is remarkably simple and geometrically intuitive: it depends only on the longest semiaxis of $\mathcal E_B$ and not on the orientation or position of the QSE. Theorem \ref{thm:canonical} is in the same vein as bounds presented in Reference \cite{PhysRevA.90.024302} that relate several other measures of quantum correlation to geometric features of QSEs.

We also note that optimal states of the form (\ref{dlc1}) have the highest quantum discord among discordant states with a given $\boldsymbol b$ that are obtained from classical states by a local trace-preserving channel. As shown in Reference \cite{PhysRevA.87.032340}, when we take a two-qubit $B$-side classical (zero discord) state and apply a channel $\Lambda_B$ to Bob's qubit, in order to create maximal $B$-side quantum discord in the output state, the optimal input state is of the form $\frac12|\psi\rangle\langle\psi|\otimes|\phi\rangle\langle\phi| +\frac12|\bar\psi\rangle\langle\bar\psi|\otimes|\bar\phi\rangle\langle\bar\phi|$, and the channel $\Lambda_B$ should have Kraus operators $E_0=|\chi\rangle\langle\phi|,E_1=|\chi'\rangle\langle\bar\phi|$, where $|\chi\rangle$ and $|\chi'\rangle$ are determined by $\boldsymbol b$.

\subsection*{Examples.}

Let us now examine some interesting classes of two-qubit states for which maximal steered coherence is easy both to find analytically and to interpret geometrically using QSEs.

\subsubsection*{X states.}
When $\boldsymbol n_B$ lies along an axis of the QSE $\mathcal E_B$ it is straightforward to see that $\mathcal C(\rho)$ is simply the length of the longest of the other two semiaxes (Figure \ref{fig:examples}\textbf{c}). All $\rho$ which are X states, i.e. have non-zero entries only in the characteristic X shape in the computational basis \cite{Yu2007}, will have such QSEs \cite{steer_njp}.

\subsubsection*{Werner states.}
As a special case of the above, when $\mathcal E_B$ is a ball of radius $r$ centered on $O'$ and $\boldsymbol n_B$ is collinear with $\overline{OO'}$, we have $\mathcal C(\rho)=r$. Furthermore, when $\mathcal E_B$ is an origin-centered ball, we have $\mathcal C(\rho)=r$ regardless of the value of $\boldsymbol b$. This allows us to evaluate $\mathcal C$ for Werner states \cite{Werner1989}, which do not in fact satisfy the non-degenerate condition $\boldsymbol b\neq \boldsymbol 0$. For a Werner state \mbox{$\rho_W=p|\Psi^-\rangle\langle\Psi^-|+\frac{1-p}{4}\sigma_0\otimes\sigma_0$} with $|\Psi^-\rangle=\frac{1}{\sqrt2}(|01\rangle-|10\rangle)$, $\mathcal E_B$ is an origin-centered ball of radius $p$ and hence $\mathcal C(\rho_W)=p$ (Figure \ref{fig:examples}\textbf{d}).

\subsubsection*{Discordant states locally created from a classical state.} We know from property (E1) that $\mathcal C$ vanishes for classical states; for a classical two-qubit state, all steered states must have the same orientation, and the QSE $\mathcal E_B$ is therefore a radial line segment. For a state obtained locally from a classical state, $\rho_\mathrm{dlc}$, the QSE is a nonradial line segment \cite{PhysRevLett.113.020402}. $\boldsymbol b$ can be any point on this segment except for the two ends of $\mathcal E_B$, which we call $\boldsymbol b_1$ and $\boldsymbol b_2$, where $b_1\geq b_2$. By definition $\mathcal C(\rho_\mathrm{dlc})$ varies for different $\boldsymbol b$; in general, we find that
\begin{equation}
b_1\sin\theta_1\leq\mathcal C(\rho_\mathrm{dlc})\bigg\{\begin{array}{ll}<b_1\sin\theta, & \mathrm{for}\ \theta\leq\frac{\pi}{2},\\
\leq b_1, & \mathrm{for}\ \theta>\frac{\pi}{2},\end{array}\label{dlc}
\end{equation}
where $\theta$ is the angle between $\boldsymbol b_1$ and $\boldsymbol b_2$ and $\theta_1$ is determined by $b_1\sin{\theta_1}=b_2\sin(\theta-\theta_1)$. From Eq. (\ref{dlc}), we see that $\mathcal C(\rho_\mathrm{dlc})$ is strictly larger than zero. In fact, $\mathcal C(\rho_\mathrm{dlc})$ can reach unity when $b_1=1$ and $\theta>\frac{\pi}{2}$.

\subsubsection*{Maximally obese states.} The general form of a maximally obese state is given by \cite{PhysRevA.90.024302}
\begin{equation}
\rho_\mathrm{mo}=\left(1-\frac{b}{2}\right)|\Psi_b\rangle\langle\Psi_b|+\frac{b}{2}|00\rangle\langle00|,
\end{equation}
where $|\Psi_b\rangle=\frac{1}{\sqrt{2-b}}(\sqrt{1-b}|01\rangle+|10\rangle)$. This is a canonical state ($\boldsymbol a = \boldsymbol 0$) with $\mathcal E_B$ centered at $(0,0,b)$ and semiaxes of length $c_1=c_2=\sqrt{1-b},c_3=1-b$ aligned with the coordinate axes. We therefore have $\mathcal C(\rho_\mathrm{mo})=\sqrt{1-b}$. It should be noted that maximally obese states maximize several measures of quantum correlation (CHSH nonlocality, singlet fraction, concurrence and negativity) over the set of all canonical states with a given marginal for Bob \cite{PhysRevA.90.024302}. Interestingly, however, they do not achieve the maximum possible $\mathcal C=\sqrt{1-b^2}$.

\section*{Discussion}

We have studied the maximal steered coherence $\mathcal C(\rho)$ for a bipartite state $\rho$. When Alice obtains a POVM outcome $M$, Bob's state is steered to $\rho_B^M$; $\mathcal C(\rho)$ is defined as the coherence of the steered state in Bob's original basis, maximized over all possible $M$. The general form of $\mathcal C$ is derived for two-qubit states. By calculating the maximal steered coherence for some important classes of two-qubit state, we find that $\mathcal C$ gives a different ordering of states from quantum entanglement or discord-like correlations. This means that $\mathcal C$ is a distinct and new measure for characterizing the remote quantum properties of bipartite states.

The maximal steered coherence vanishes only when $\rho$ is a classical state, and $\mathcal C(\rho)$ can be increased by local trace-preserving channels. For a two-qubit state $\rho$ we derive a necessary and sufficient condition for a local qubit channel to be capable of increasing $\mathcal C(\rho)$. This is in fact identical to the condition for increasing quantum discord, suggesting that local increase of quantum discord might be used in a protocol for increasing steered coherence.

Finally, we consider the relevance of $\mathcal C$ from a more physical perspective by presenting a concrete example in which steered coherence can be exploited. Say that Alice and Bob share a two-qubit state of the form \eqref{dlc1} with $|\psi\rangle=|0\rangle$, $|\bar\psi\rangle=|1\rangle$ and $\boldsymbol b=(0,0,b)$ (which, as illustrated in Figure \ref{fig:examples}\textbf{b}, will lie at the midpoint of the chord $\mathcal E_B$ joining $|\chi\rangle$ and $|\chi'\rangle$). Suppose also that Alice's and Bob's systems are described by the local Hamiltonian $H=-\hbar \omega\sigma_3$. Let us restrict Alice's and Bob's local operations to those which are covariant with respect to time-translation symmetry~\cite{marvian2014extending}. For Alice, these operations are the ones for which $\Phi\left(e^{-iHt}\rho_A e^{iH t}\right)=e^{-iHt}\Phi(\rho_A)e^{iHt}$; and similarly for Bob. Physically, this restriction corresponds to local energy-conserving unitaries with the assistance of incoherent environmental ancillas~\cite{keyl1999optimal}: Alice's operation is covariant if and only if it can be written as $\Phi(\rho_A)=\mathrm{tr}_E[U (\rho_A\otimes\xi_E)U^\dag],$ where $U$ is a unitary, $H_E$ is the Hamiltonian of the ancilla, $[U,H+H_E]=0$ and $[\xi_E, H_E]=0$; and similarly for Bob. The set of covariant operations is a strict subset of incoherent operations~\cite{PhysRevLett.113.140401} and a strict superset of thermal operations~\cite{brandao2011resource}.

Bob's reduced state $\rho_B=\frac{1}{2}(\mathbb{1}+b \sigma_3)$ is incoherent in his energy eigenbasis $\{|0\rangle, |1\rangle\}$, and his local covariant operations alone cannot generate any coherence. However, by performing a $\sigma_3$ measurement, which is a covariant operation, and classically communicating the result to Bob, Alice steers him to either $|\chi\rangle$ or $|\chi'\rangle$, states that are manifestly coherent in the energy eigenbasis. $\mathcal C(\rho)$ gives a measure of the maximal coherence that Alice can induce on Bob's system by steering. In this way, Alice remotely `activates' a coherent state for Bob that he was unable to produce himself. Bob may now use this coherence as a resource for quantum information processing tasks, e.g. work extraction by a thermal machine, which is known to be enhanced in the presence of a coherent reference system~\cite{workextraction}. Given the ever-increasing number of applications for coherence found throughout quantum information science, one can envisage a range of such scenarios in which steered coherence could be used as a resource.

\section*{Methods}

\subsection*{Proof of property (E1).}
The `if' part is obvious: Bob's reduced state is $\rho_B=\sum_{i}p_i|\xi_i\rangle\langle\xi_i|$, and the steered state is $\rho_B^M=\sum_ip_i\mathrm{tr}(\rho_i^A M)|\xi_i\rangle\langle\xi_i|$. These are both diagonal in the basis $\{|\xi_i\rangle\}$, and hence $\mathcal C(\rho)=0$.

For the `only if' part, first consider a separable state $\rho_s=\sum_ip_i\rho_i^A\otimes\rho_i^B$. When $\mathcal C(\rho_s)=0$, the steered states $\rho_B^M=\sum_ip_i\mathrm{tr}(\rho^A_i M)\rho_i^B$ for different POVM operators $M$ should commute with each other, which is equivalent to all $\rho_i^B$ commuting with each other. So $\mathcal C(\rho_s)$ vanishes only if it is in the form (\ref{classical}). For an entangled state $\rho_e$, we express $\rho_e$ in the optimal pure state decomposition form as $\rho_e=\sum_ip_i|\Phi_i\rangle\langle\Phi_i|$, so that the entanglement of formation is $E_F(\rho_e)=\sum_ip_iE_F(|\Phi_i\rangle\langle\Phi_i|)$. Since $\rho_e$ is entangled, at least one of the $|\Phi_i\rangle$ is entangled. Hence, for $\rho_e$, it is not possible for all of Bob's steered states to share the same eigenbasis; this means that $\mathcal C(\rho_e)\neq0$ for any entangled $\rho_e$.

\subsection*{Proof of Theorem \ref{thm:increase}.}
A channel $\Lambda_B$ that is neither unital nor semi-classical can increase $\mathcal C$, because such channels can transform a classical state with vanishing $\mathcal C$ into a discordant state with nonzero $\mathcal C$ \cite{PhysRevLett.107.170502,PhysRevA.85.032102}. We now focus on the `only if' part, and prove that a local unital channel or a local semi-classical channel cannot increase $\mathcal C$ for any two-qubit input state.

A semi-classical channel $\Lambda^{sc}$ \cite{PhysRevLett.107.170502}, which maps any input state $\rho$ to a state with zero coherence in a given basis $\Lambda^{sc}(\rho)=\sum_kp_k(\rho)|k\rangle\langle k|$, yields $\mathcal C=0$ for any input state. As proved by King and Ruskai \cite{222939961}, any unital channel is equivalent to $\Lambda^u(\cdot)=\sum_{i=0}^3e_i\sigma_i(\cdot)\sigma_i$,
where $0\leq e_i\leq1$ and $\sum_i e_i=1$. The effect of this channel on a qubit state is to shrink the Bloch vector as $\Lambda^u:\ \boldsymbol b\rightarrow\boldsymbol b'=(p_1b_1,p_2b_2,p_3b_3)^{T}$, where $p_1=e_0+e_1-e_2-e_3$, and $p_{2,3}$ are related to $e_i$ in a similar way. Let $\boldsymbol b_M$ be a steered state for the input state $\rho$. Then the coherence of $\boldsymbol b_M$ is $C(\boldsymbol b_M,\boldsymbol b)=|\boldsymbol b_M\times\boldsymbol b|/b$. Under the action of $\Lambda^u$, the steered state and Bob's reduced state become $\boldsymbol b_M'$ and $\boldsymbol b'$ respectively, and the coherence of $\boldsymbol b_M'$ in the eigenbasis of $\boldsymbol b'$ is
\begin{eqnarray}
C^2(\boldsymbol b_M',\boldsymbol b')=\frac{(b_{M1}b_3-b_{M3}b_1)^2p_1^2p_3^2}{(b_1p_1)^2+(b_2p_2)^2+(b_3p_3)^2}
+(1\rightarrow2\rightarrow3)+(1\rightarrow3\rightarrow2).
\end{eqnarray}
If the inequality
\begin{equation}\label{ineq}C(\boldsymbol b_M',\boldsymbol b')\leq C(\boldsymbol b_M,\boldsymbol b)\end{equation}
holds then the maximal steered coherence for the output state $\mathcal C(\iden_A\otimes\Lambda_B^u(\rho))=C(\boldsymbol b_{M_\mathrm{opt}}',\boldsymbol b')\leq C(\boldsymbol b_{M_\mathrm{opt}},\boldsymbol b)\leq\mathcal C(\rho)$, where $M_\mathrm{opt}$ is the optimal POVM operator to maximize (\ref{coherence}) for the output state and $\boldsymbol b_{M_\mathrm{opt}}$ is the corresponding input state for $\boldsymbol b_{M_\mathrm{opt}}'$. Hence it is sufficient to prove that \eqref{ineq} holds for some $\boldsymbol b_M$ and $\boldsymbol b$. Note that $\frac{(b_{M1}b_3-b_{M3}b_1)^2p_1^2p_3^2}{(b_1p_1)^2+(b_3p_3)^2}\leq\frac{(b_{M1}b_3-b_{M3}b_1)^2}{b_1^2+b_3^2}$. By using the fact that $\frac{A+B}{C+D}<\frac{A'+B}{C'+D}$ for $0<B<D,0<A<C,0<A'<C'$ and $\frac{A}{C}>\frac{A'}{C'}$, we arrive at $C^2(\boldsymbol b_M',\boldsymbol b')\leq C^2(\boldsymbol b_M,\boldsymbol b)$, which is equivalent to (\ref{ineq}).

\subsection*{Proof of Theorem \ref{thm:canonical}.}
The steered state $\boldsymbol b_M$ which achieves the maximum in Eq. (\ref{C_eqn}) corresponds to a point $B_M$ on the surface of $\mathcal E_B$. We have $\mathcal C(\rho_\mathrm{can})=C(\boldsymbol b_M,\boldsymbol n_B)=D(B_M,\overline{OB})\leq\overline{BB_M}\leq c_1$, where $D(B_M,\overline{OB})$ is the perpendicular distance between $B_M$ and $\overline{OB}$. To ensure that $\mathcal E_B$ lies inside the Bloch sphere we require that $c_1\leq\sqrt{1-b^2}$.

To saturate the bound we take $c_1=\sqrt{1-b^2}$, but we must also demonstrate that $c_2=c_3=0$, i.e. that $\mathcal E_B$ cannot be an ellipsoid or an ellipse. We know that $\mathcal E_B$ must meet the Bloch sphere at two points, corresponding to the pure states $|\chi\rangle$ and $|\chi'\rangle$. Firstly suppose that $\mathcal E_B$ is a three-dimensional ellipsoid. Elementary geometry tells us that the surface of an ellipsoid at the end of any axis must be perpendicular to that axis. The points at the ends of the $c_1$ axis on $\mathcal E_B$ lie on the surface of the Bloch sphere. Since the surface of $\mathcal E_B$ must lie perpendicular to the $c_1$ axis at these points, $\mathcal E_B$ must puncture the surface of the Bloch sphere. Such $\mathcal E_B$ cannot represent a physical two-qubit state and so $\mathcal E_B$ cannot be an ellipsoid. Now consider the case that $\mathcal E_B$ is an ellipse. The nested tetrahedron condition tells us that any degenerate $\mathcal E_B$ describing a physical state must fit inside a triangle inside the Bloch sphere~\cite{PhysRevLett.113.020402,steer_njp}. Geometrically, no ellipse that touches the Bloch sphere at two points can satisfy this, and so $\mathcal E_B$ cannot be an ellipse. $\mathcal E_B$ must therefore be a line, i.e. the chord going between $|\chi\rangle$ and $|\chi'\rangle$; this corresponds to the state (\ref{dlc1}).

\bibliography{sample}

\section*{Acknowledgements}

This work was supported by NSFC under Grant No. 11447161,11504205, the Fundamental Research Funds of Shandong University under Grant No. 2014TB018, and the National Key Basic Research Program of China under Grant No. 2015CB921003. A.M. is funded by EPSRC. We thank Kamil Korzekwa, Matteo Lostaglio and Terry Rudolph for useful discussions.

\section*{Author contributions statement}

X.H. contributed the idea. X.H., A.M. and B.Z. performed the calculations. X.H., A.M. and H.F. wrote the paper. All authors reviewed the manuscript and agreed with the submission.

\section*{Additional information}

\textbf{Competing financial interests:} The authors declare that they have no competing financial interests.

\end{document}